# Freeform microfluidic networks encapsulated in laser printed three-dimensional macro-scale glass objects

Zijie Lin[1,2], Jian Xu[1,2*], Yunpeng Song[1,2], Xiaolong Li[1,2],

Peng Wang[2], Wei Chu[2], Zhenhua Wang[1,2] & Ya Cheng[1,2,3,4*]

[1]State Key Laboratory of Precision Spectroscopy, School of Physics and Electronic Science, East China Normal University, Shanghai 200062, China

[2]XXL—The Extreme Optoelectromechanics Laboratory, School of Physics and Electronic Science, East China Normal University, Shanghai 200241, China

[3]Collaborative Innovation Center of Extreme Optics, Shanxi University, Taiyuan, Shanxi 030006, China

[4]Collaborative Innovation Center of Light Manipulations and Applications, Shandong Normal University, Jinan 250358, China

[*]Correspondence and requests for materials should be addressed to J. X. (email: jxu@phy.ecnu.edu.cn) or Y. C. (email: ya.cheng@siom.ac.cn)

**Abstract:**

Large-scale microfluidic microsystems with complex three-dimensional (3D) configurations are highly in demand by both fundamental research and industrial application, holding the potentials for fostering a wide range of innovative applications such as lab-on-a-chip and organ-on-a-chip as well as continuous-flow manufacturing of fine chemicals. However, freeform fabrication of such systems remains challenging for most of the current fabrication techniques in terms of fabrication resolution, flexibility, and achievable footprint size. Here, we report ultrashort pulse laser microfabrication of freeform microfluidic circuits with high aspect ratios and tunable diameters embedded in 3D printed glass objects. We achieve uniform microfluidic channel diameter by carefully distributing a string of extra access ports along the microfluidic channels for avoiding the over-etching in the thin microfluidic channels. After the chemical etching is completed, the extra access ports are sealed using carbon dioxide laser induced localized glass melting. We demonstrate a model hand of fused silica with a size of ~3 cm × 2.7 cm × 1.1 cm in which the whole blood vessel system is encapsulated.

In the past decades, microfluidic technology has emerged as a powerful methodology for applications in many fields such as lab on chips, organs on chips, biopharmaceuticals, and fine chemical industries [1-6]. As a core element of microfluidic technology, high-performance fabrication of hollow microchannels with freeform geometries is crucial for further developing innovative microfluidic technology. In particular, extension of the microfluidic networks from widely used two-dimensional (2D) to three-dimensional (3D) configurations has now been considered as a promising scheme to enhance the performance of manipulation of fluids such as high-efficiency mixing, separation and detection [7-11].

Fused silica is one of the widely used substrates for the microfluidic technology in laboratory research and industrial applications due to its high melting point, high chemical stability, low thermal expansion coefficient, wide transmission spectral range, and good biocompatibility. However, fabrication of freeform 3D hollow microchannels in fused silica is still difficult for current 2D fabrication techniques in terms of complex and tedious fabrication procedures, additional costs for aligning, stacking, and bonding steps [12-14]. Currently, one of the most representative 3D microfabrication techniques of fused silica is ultrashort pulse laser microfabrication [15-20]. A suspended hollow microchannel structure with a 3D controllable configuration can be fabricated in a way of either laser assisted selective wet etching [21-23] or liquid-assisted laser ablation [24-26]. In addition, suspended hollow structures with 3D shapes and a high precision in fused silica can be fabricated by combining room-temperature casting and sacrificial template replication [27]. Meanwhile, 3D printing of glass materials emerged in recent years brings some new opportunities for scalable fabrication of freeform glass structures by either additive manufacturing strategies [28-33] or laser subtractive processing methods [34-36]. However, controllable formation of freeform hollow microchannel with arbitrary length encapsulated in 3D printed glass structures has not yet been demonstrated, which is highly desirable for high-infidelity manufacturing of artificial organs, on-chip construction of biomimetic microenvironments [37-41], as well as enrichment of the functions of continuous-flow microreactors (e.g., seamless integration of customized functional module for temperature control and on-line spectrometer monitoring) [42, 43]. Especially, simultaneous fabrication of 3D embedded hollow microchannels with high aspect ratios and centimeter-scale glass objects with external arbitrary shapes still remains challenging for current microfabrication technologies in terms of fabrication resolution, flexibility, and achievable footprint size. For instance, to form encapsulated microchannel networks during stereolithography 3D printing of glass objects, it is difficult to fully remove the entrapped unexposed material and accurately control the material's exposure inside 3D microvoids for avoiding unwanted blocking of channels.

Herein, we demonstrate fabrication of freeform 3D microfluidic networks encapsulated in 3D printed glass macro-scale objects using the hybrid laser microfabrication scheme, which based on the combination of ultrashort pulse laser assisted chemical etching of glass, 3D laser subtractive glass printing and carbon oxide ($CO_2$) laser induced glass melting. To fabricate 3D uniform glass microchannels with centimeter-scale length, introduction of a string of extra-access ports along the microfluidic channels has been adopted for ensuring homogeneous fabrication of the channels during the chemical etching [44, 45]. Then, controllable sealing of those ports has been achieved using $CO_2$ laser irradiation in a defocusing manner for all-glass structure manufacturing [46-48]. Moreover, 3D laser subtractive glass printing enables rapid manufacturing of macro-scale glass objects with desirable shapes and a high precision down to several tens of microns due to unique polarization-insensitive and depth-insensitive processing characteristics of picosecond laser irradiation [34, 49]. The proposed approach allows both freeform fabrication of 3D encapsulated microchannels and 3D printable glass structures with arbitrary lengths and flexible configurations, paving the way for advanced manufacturing of 3D large-scale all-glass microfluidic systems.

# Results

## $CO_2$ laser defocusing irradiation of extra-access ports of glass channels

To prepare 3D microchannels with arbitrary lengths inside fused silica, the fabrication procedure consists of three main steps as illustrated in Fig.1a. Firstly, micropatterns of 3D microchannels including main channels and a string of extra-access ports were drawn inside glass by ultrashort pulse laser direct writing (see Fig. 1b). Secondly, the laser-irradiated glass samples were immersed in an ultrasonic bath of KOH solution (10 mol/L) at ~85 °C until the modified regions including the micropatterns of microchannels and extra-access ports were fully removed from glass matrix. Finally, the chemically etched glass samples were selectively melted for sealing of extra-access ports using a carbon dioxide laser as illustrated in Fig. 1c. To obtain the microchannel with uniform diameters, introduction of a string of extra-access ports which connected to the microchannels is adopted for avoiding the over-etching in the channels [44, 45]. To seal the extra-access ports using $CO_2$ laser irradiation, it is necessary to control the irradiation process to drive the sufficient molten glass around irradiated zones immigrate from the peripheral area to the openings of the ports. Therefore, observation and determination of the dimensions of laser-melted areas (laser affected zones) on pristine fused silica surface under different laser powers, irradiation durations and defocusing distances (see Supplementary Figure 1) has been performed. With the increase of laser power from 9 W to 30 W, the dimension of the laser-melted area on glass surface increases in each condition (see

Supplementary Figure 2a). Moreover, when $CO_2$ laser irradiation was defocusing from 0 cm (i.e., direct focusing on glass surface) to 5 cm, the dimension of the laser-melted area on glass surface also increases at the same condition. In comparison, by prolonging the irradiation duration from 20 s to 60 s, the dimension of laser-melted area slightly increases in each condition.

To evaluate the dimension of a laser-melted area on glass surface, herein a diameter of a surface crater induced by laser irradiation is defined as a critical parameter) (see the inset of Supplementary Figure 2b). With the increase of laser power, the diameter of the surface crater increases much faster at a defocusing distance of 5 cm than that of 0 cm (i.e., direct focusing of $CO_2$ laser beam on glass surface) (see Supplementary Figure 2b). Especially, the average diameter of the generated surface crater under defocusing irradiation of $CO_2$ laser beam at 5 cm can reach ~1426 μm when the laser power and irradiation duration are set at 30 W and 60 s, respectively, which is about ~1.5 times that of direct focusing on glass surface (~931 μm), indicating the defocusing scheme is an effective approach for controlling the melted area of $CO_2$ laser irradiation on fused silica surface. It should be pointed out that although the diameter of the surface crater increases at the same condition using the defocusing scheme, the depth of the surface crater decrease as compared with direct focusing. The average depth of the generated surface crater under defocusing irradiation of $CO_2$ laser beam at 5 cm at a laser power of 30 W for 60 s is about ~1750 μm, which is only ~0.63 times that of direct focusing on glass surface (~2761 μm, see Supplementary Figure 3).

To apply $CO_2$ laser induced localized melting for the sealing of extra-access ports of microchannels, a series of glass microchannels with different diameters of extra-access ports ranging from 150 μm to 350 μm in fused silica using ultrashort pulse laser assisted chemical etching have been prepared. As shown in Fig. 2a, all extra-access ports can be fully sealed using $CO_2$ laser defocusing irradiation with the distances of 10 cm and 12.5 cm at a laser power of 33 W and an irradiation duration of 60 s. One can clearly see that surface craters on glass substrates and sealing layers which block the extra-access ports can simultaneously formed after $CO_2$ laser induced melting.

To evaluate the performance and robustness of the sealing process, the thickness and the depth of the sealing layer are defined as two critical parameters, which are illustrated in the insets of Fig 2b and Fig. 2c, respectively. In the case of defocusing distance at 10 cm (the middle panel of Fig. 2a), the thickness of the sealing layer is almost beyond 200 μm for sealing the ports with the diameters of 150 μm, 200 μm and 250 μm. However, with the further increase of the diameter of the port (e.g., 300 μm and 350 μm), the thickness of the sealing layer decreases dramatically. The main reason is that the molten volume of glass induced by $CO_2$ laser irradiation is limited by laser-melted area, which is not enough to form a thick sealing layer as the volume of the port increases. To form a thick sealing layer of a port with a large diameter, further increase of defocusing distance is a convenient way for the increase of the laser-melted volume. When the defocusing distance increase to 12.5 cm (see the right panel of Fig 2a), the thickness of the sealing layer in each condition can be over 300 μm, indicating the stability of the sealing process enabled by $CO_2$ laser defocusing irradiation.

Besides the thickness of the sealing layer, the depth of the sealing layer which is correlated to the designed depth of the port is also important parameter. To minimize dead-end volumes and improve the performance of glass microchannels with sealed extra-access ports for microfluidic applications, it is necessary to control the sealing process for minimizing the distance between the sealing layer and the microchannel. The depth of the sealing layer is defined as the distance between the top of the sealing layer and the laser-melted glass surface, which can be controlled by tuning the defocusing distance. As presented in Fig. 2c, the depth of the sealing layer at defocusing distance of 12.5 cm is smaller than that of 10 cm in each condition due to the increase of laser-induced melted area. Moreover, with the increase of the diameter of the port, the depth of the layer at 10 cm increase rapidly. The possible reason is that to fully seal the port with a large diameter, when melted volumes of glass are relatively limited, the availably immigrated molted glass may need to continuously move downwards driven by the surface tension until the whole sealing stabilizes, leading to the increase of the depth of the final sealing layer. In contrast, the depth of the layer at 12.5 cm is more stable due to its abundant melted area. In general, a successful sealing of extra-access ports has to meet two critical conditions. One is that the sealing of the ports with different diameters have to be well controlled by laser processing conditions. Another is that the microchannel should be not interrupted and blocked during the whole sealing process. Therefore, to ensure safe and robust sealing of a microchannel, normally a designed depth of an extra-access port should be not less than that of the final sealing layer (see Supplementary Tables 1 and 2).

As for the possible mechanism of $CO_2$ laser induced sealing of the ports, we infer that the sealing processes may consist of several vital stages as follows (see Supplementary Figure 4a). When an extra-access port is initially illuminated by defocusing $CO_2$ laser beam, the temperature around the peripheral area of the port rapidly increase due to the effective absorption of laser energy. When the temperature is close to the melting point of the glass matrix, the surface area around the port starts to become a portion of gathered liquid melts which acts as a flowable media under the continuous laser irradiation. The flowable fluid tends to immigrate towards the center of the port driven by the surface tension, resulting in the shrinkage of the port. With the increase of irradiation durations at a certain laser influence, the center of port continuously decreases, and finally the port can be blocked with a formation of a sealing layer. After that, the sealed layer tends to be saturated under continuous in-situ heating of $CO_2$ laser beam. Meanwhile, the surface crater simultaneously forms on glass surface as the sealing layer forms below the glass surfaces.

To identify aforementioned assumption, the influence of different irradiation durations on the sealing process of the extra-access ports was performed. With $CO_2$ laser irradiation with a defocusing distance of 12.5 cm, the extra-access ports with diameters of ~300 μm at different irradiation durations exhibit different sealing results (see Supplementary Figure 4b). For the irradiation time of 5 s, the shape of an extra-access port was almost no change under the laser irradiation. For the longer irradiation durations ranging from 10 s to 35 s, the shapes of the ports change clearly due to in-situ melting and immigration of glass around the peripheral areas of ports under $CO_2$ laser irradiation. Moreover, with prolonging the irradiation duration the shape of the ports exhibits a clear tendency of shrinkage towards the central area of the ports driven by the laser actions. For the irradiation time of 40 s, the port is nearly blocked except for a small portion of the central part, which to some extent corresponds a near-critical condition between sealing and unsealing. And the port can be fully sealed for the irradiation time of 45 s. For further prolonging irradiation duration (e.g., 60 s), the sealing process becomes stabilized due to the balance between the surface tension and laser induced thermally driven force. In short, the experimental result is consistent with aforementioned explanation of the sealing mechanism as well as the results of Fig. 2.

## Freeform fabrication of 3D channels with arbitrary configurations and lengths

To demonstrate the applicability of the proposed approach for fabrication of 3D microchannels with flexible configurations, a helical microchannel with a set of extra-access ports has been prepared by ultrashort pulse laser assisted chemical etching as shown in Fig. 3a. With the introduction of the ports distributed along the helical microchannel, the uniformity of the microchannel can be easily identified. By use of $CO_2$ laser induced melting, all the extra-access ports can be fully sealed (Fig. 3b), indicating the stability and versatility of the sealing process. Further, we fabricated two 3D intertwined microchannels with extra-access ports as shown in the inset of Fig. 3c. After sequenced $CO_2$ laser irradiation on each extra-access port, all the extra-access ports can be also fully sealed (Fig. 4d). Further microfluidic experiments of the intertwined microchannels with sealed extra-access ports confirmed that by filling different color (red and purple) solutions in the channels both channels were successfully throughout without any leakage and each channel remains individually isolated without disturbances, indicating a high-precision sealing ability using the proposed approach.

Besides the arbitrary configuration, in principle, the fabricated length of the microchannel can be almost unlimited as long as the processing distance of the stage permits. For instance, a 7 cm length straight channel with a set of sealed extra-access ports can be easily fabricated inside a glass chip (see Supplementary Figure 5). In addition, this approach can be also used for sealing the spiral microchannel deeply embedded inside glass substrate (see Supplementary Figure 6).

## 3D laser subtractive printing of scalable glass hands

Besides its powerful capability for versatile fabrication of 3D microchannels in glass, ultrashort pulse laser assisted chemical etching can be applied for 3D subtractive printing of macro-scale glass structures with arbitrary geometries and high precision. As we mentioned before, 3D laser subtractive printing adopted in this work relies on the picosecond laser irradiation. It is well known that when the fused silica is irradiated with focused femtosecond laser beams, space selective chemical etching can be achieved [35, 50-52]. The etching rate depends sensitively on the polarization of the laser beam. However, by chirping the Fourier-transform-limited femtosecond laser pulses to picosecond pulses, the polarization dependence of the etching rate surprisingly disappeared due to the elimination of nanograting structures, whereas an efficient etching rate could still be maintained [49]. Such a polarization-insensitive selective etching behavior in fused silica under picosecond laser irradiation in the regime between 4 ps and 10 ps has a significant implication for applications such as high-precision and 3D isotropic glass printing and large-scale fabrication of 3D microfluidic systems. In addition, we have found that the depth-insensitive focusing of picosecond pulses at optimum pulse energy allows the laser structuring can maintain an unchanged feature size from the bottom to the top of the glass without the additional compensation of the aberration originated from the refractive index mismatch between air and glass, enabling laser subtractive printing of centimeter-scale 3D glass structures at a micro-scale feature size in a flexible manner [34, 53]. With ultrashort pulse laser 3D subtractive printing, 3D arbitrary glass objects at a centimeter scale can be easily produced. As shown in Fig. 4, the back and the palm of three glass hands with scalable footprint sizes can be fabricated by 4 ps laser rapid structuring and selective etching of the reverse mold patterns of the 3D hand.

**Biomimic fabrication of a vascular structure in a 3D printed hand**

To combine both the advantages of 3D fabrication of microfluidic channels and high-precision printing of glass, a bioinspired blood-vessel structure encapsulated in a 3D printed glass hand has been designed and successfully fabricated. As illustrated in Figs. 5a and 5b, a 3D hand model and an embedded micropattern were first individually designed and further fused together for ultrashort pulse laser assisted chemical etching. An embedded micropattern including a microchannel and several tens of extra-access ports was first created inside glass using the ultrashort pulse laser direct writing, and then the surrounding area of the hand model was modified by the same laser beam in a 3D programmable manner. After that, the laser-modified glass sample was chemical etched until all laser-irradiated regions were fully removed. Figure 5c presents a fabricated hand model with a whole size of about ~ 3 cm × 2.7 cm × 1.1 cm, in which fingernails and some skin texture patterns on the hand back can be clear distinguished due to the high precision of 3D subtractive printing down to several tens of micrometers enabled by ultrashort pulse laser assisted chemical etching (see Supplementary Figure 7). Moreover, the embedded microchannel patterns including open extra-access ports and three openings below the hand can be observed. Finally, all extra-access ports were sealed by $CO_2$ laser defocusing irradiation to form a 3D close microchannel encapsulated in the hand model with only three openings as shown in Fig. 5d. Further microfluidic experiments confirmed that the encapsulated microchannel with sealed extra-access ports are fully throughout without any leakage (see Fig. 5e). Especially, the microfluidic circuit can be clearly visualized inside the close-up image of fingers as indicated by a white arrow in Figure 5e (see Fig. 5f). We expect that this approach based on combination of ultrashort pulse laser assisted chemical etching, 3D laser subtractive printing and $CO_2$ laser induced melting for biomimetic fabrication of visible vascular structures in 3D printed glass structures will be beneficial for development of novel lab-on-a-chip and biochip devices to promote the understanding causes of human diseases such as cancers, cardiovascular diseases.

## Discussion

A simple, robust and reproducible approach based on ultrashort pulse laser assisted chemical etching of glass and $CO_2$ laser induced localized sealing of glass ports have been proposed for large-scale fabrication of freeform all-glass microfluidic networks encapsulated in 3D printed macro-scale objects. As compared with conventional fabrication methods, the proposed methods have several distinguished advantages as follow. First, it provides a powerful processing capability for both high-precision (down to several tens of micrometers) and large-scale (up to several centimeters) production of freeform suspended glass microchannels in 3D printed glass objects. Both polarization-insensitive and depth-insensitive processing characteristic enabled by picosecond laser structuring provides a unique capability for accomplishment of simultaneous fabrication of embedded freeform microfluidic networks and laser printed macro-scale glass 3D objects at a resolution of several tens of micrometers in simple and flexible manner. The introduction of extra-access ports improves the performance of selective chemical etching and break the length limits of fabricating 3D homogeneous channels. In turn, the reduction of etching time also enables high precision and high quality of fabrication of glass microchannels. Second, it enables controllable sealing of extra-access ports for all-glass microfluidic applications. The molten glass can be formed in laser irradiation area due to the high temperature induced by $CO_2$ laser heating and then driven by surface tensions to immigrate the extra-access ports for formation of closed channels with only few ports. The dynamical process of sealing can be well controlled and a processing window for crack-free and high-precision sealing can be reliably obtained by adjusting the laser power, irradiation duration, and focusing conditions. Last but not least, all-glass solutions of this approach inherently exhibit superior chemical inertness for practical microfluidic applications of non-alkaline solutions. As compared with a PDMS sealing method which often are sealed on the surface of chip substrates, this approach allows robust formation of a sealing layer with tunable depths and controllable thickness, which exhibits superior mechanical strengths for long-term use in harsh environments. It should be noted that both the roughness of the inner walls of microchannels and the printed glass surfaces can be further decreased since the ultrashort pulse laser fabricated microchannel after an optimum heat treatment process can be used for optical waveguiding application due to great improvement of surface smoothness (e.g., the total transmission of the hollow waveguide reaches ~90%) [44]. Moreover, further integration of photonic components such as optical waveguides, microgratings and microelectric elements will enrich the functionality of 3D laser-manufactured microfluidic devices and systems in near future.

## Methods

**Preparation of 3D microchannels with arbitrary lengths**

In all experiments, fused silica glass samples (Corning 7980) with six-side polished surfaces and different sizes were used as processing substrates. An ultrashort laser system (Light Conversion, Pharos 20 W) with a repetition rate of 250 kHz, a variable pulse duration ranging from 270 fs to 10 ps was used as 3D laser modification. The laser beam was focused by an objective lens with a numerical aperture of 0.3 and a transmission rate of ~50% at 1030 nm. To prepare the glass microchannels, the pulse energy and the writing speed were set at ~2.4 μJ and 50 mm/s, respectively. Two pulse durations (270 fs and 4 ps) were chosen in most experiments depending on the specific requirement. For 270 fs, the chemical etching of channel exhibits a polarization-dependence tendency [50-52]. For 4 ps, the etching of channel is polarization-insensitive with a rapid etching rate as we previously reported [49]. The fused silica samples were amounted in a 3D programmable air-bearing stage (Aerotech, X-Y-Z stage) for spatially selective laser modification. A lamp was used to illuminate the glass samples from the bottom and a CCD camera was employed to monitor the whole processing in real time. After laser direct writing, the glass samples were immersed in an ultrasonic bath of KOH solution (10 mol/L) at ~85 °C for chemical etching until the modified regions including the micropatterns of microchannels and extra-access ports were fully removed. Finally, the chemically etched glass samples were selectively melted for sealing of extra-access ports using a carbon dioxide laser (Synrad, Firestar OEM vi30). The repetition rate of the laser beam was set at 10 kHz. As depicted in the inset of Fig. 1c, the openings of extra-access ports on glass surfaces can be localized melted and blocked by focusing the $CO_2$ laser beam with a ZnSe lens with a focal length of 200 mm, which leads to form the closed microchannel with only two ports. For defocusing irradiation of the $CO_2$ laser beam (see Supplementary Figure 1), the defocusing distances ranging from 0 cm to 15 cm were controlled by a manual linear stage.

**3D laser subtractive printing for 3D glass hands with an encapsulated microfluidic circuit**

For design and programming of a 3D model, a model pattern of a 3D freeform hand was first constructed and then transformed to the reverse mold patterns of the hand with a format of stereolithography (STL) file using a commercial graphic design software. Furthermore, the STL file of the reverse mold patterns was sliced using a 3D printing software and further converted to a G-code file for subsequent implementation of laser processing. In addition, command codes of the electronic shutter were incorporated into the aforementioned G-code file for preciously controlling the ON and OFF state of light beam during laser direct writing. For ultrashort pulse laser direct writing, the programmed microfluidic circuit micropatterns were first created by the focused ultrashort pulse laser beam in a fused silica glass with size of 4 cm × 4 cm × 1.2 cm for several minutes, and the reverse mold patterns of the 3D hand were written by the same laser for ~20 hours. To create the microfluidic circuit micropatterns including a main microchannel and a set of extra-access ports (50 ports) with a certain spacing ranging from 3 to 5 mm, the ultrashort laser direct writing processing parameters were chosen as follows: pulse duration: 4 ps, repetition rate: 250 kHz, central wavelength: 1030 nm, writing speed: 40 mm/s, pulse energy before objective lens: 4.08 μJ, numerical aperture of objective lens: 0.3.

To write the reverse mold patterns of the 3D hand, the ultrashort laser direct writing processing parameters were chosen as follows: pulse duration: 4 ps, repetition rate: 500 kHz, central wavelength: 1030 nm, writing speed: 80 mm/s, pulse energy before objective lens: 2.26–2.96 μJ (depending on the specific processing depth), numerical aperture of objective lens: 0.3. For selective chemical etching, the laser treated glass samples were placed in an ultrasonic bath first with 5% HF solution in room temperature for 8 hours and then with a 10 mol/L KOH solution at 85 ℃ for several tens of hours until the whole modified region was fully removed. To obtain the fully through microfluidic channels and extra-access ports, the selective chemical etching was taken about several days. For $CO_2$ laser defocusing irradiation, the extra-access ports were sealed using $CO_2$ laser irradiation one by one with a defocusing distance of 12.5 cm. The laser power for sealing the ports around the fingers and the back of the hand were set at 30 W and 36 W, respectively.

## Characterization

The morphologies of processing glass samples were observed and recorded by an optical microscopy (Olympus, BX53) and a digital camera (Canon, EOS 6D).

**Acknowledgments**

This work has been funded by the Key Project of Science and Technology Commission of Shanghai Municipality (Grant 18DZ1112700).

## Author contributions

Y. C. conceived the concept. J. X. and Y. C. designed and supervised the experiments. Z. L., J. X., and Y. S. performed the experiments. Z. L., J. X. and Y. C. analyzed and processed the data. X. L. and Z. W. contributed to laser assisted chemical etching of glass and $CO_2$ laser processing. P. W. and W. C. contributed to electronic control on laser subtractive printing. J. X., Z. L. and Y. C. wrote the paper.

## Competing interests

The authors declare no competing interests.

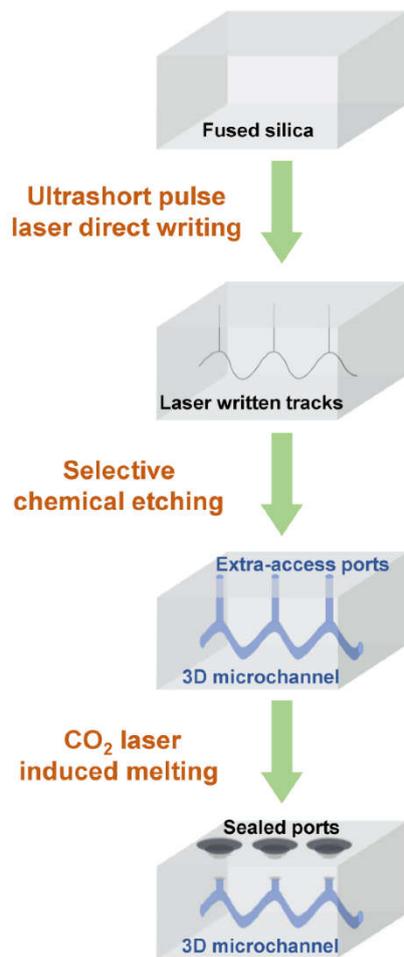
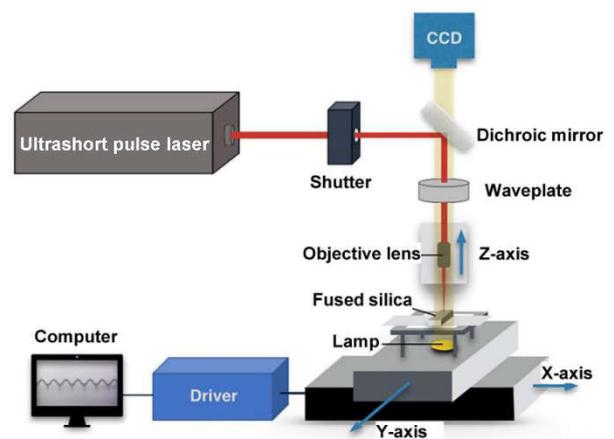
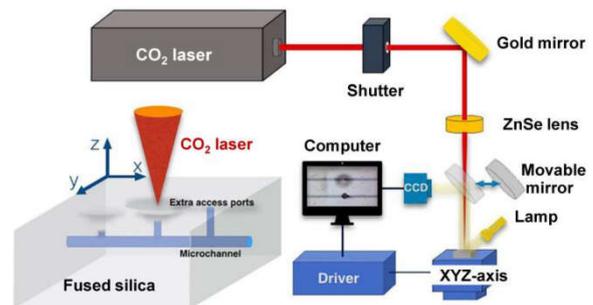

**Figure 1.** Schematic of fabrication procedure for 3D microchannel with arbitrary length based on hybrid laser microfabrication, which consists of three main steps: Ultrashort pulse laser direct writing of patterns including microchannels and extra-access ports in fused silica, selective chemical etching of glass, and $CO_2$ laser induced melting for sealing of the extra-access ports. Schematics of experimental setups for (b) 3D processing of ultrashort pulse laser modified patterns of microchannels with extra-access ports in fused silica glass and (c) localized sealing of extra-access ports of fabricated glass channels using $CO_2$ laser induced melting. Inset in (c) indicates that the extra-access ports can be sealed by $CO_2$ laser in sequence to form a closed microchannel with only two ports.

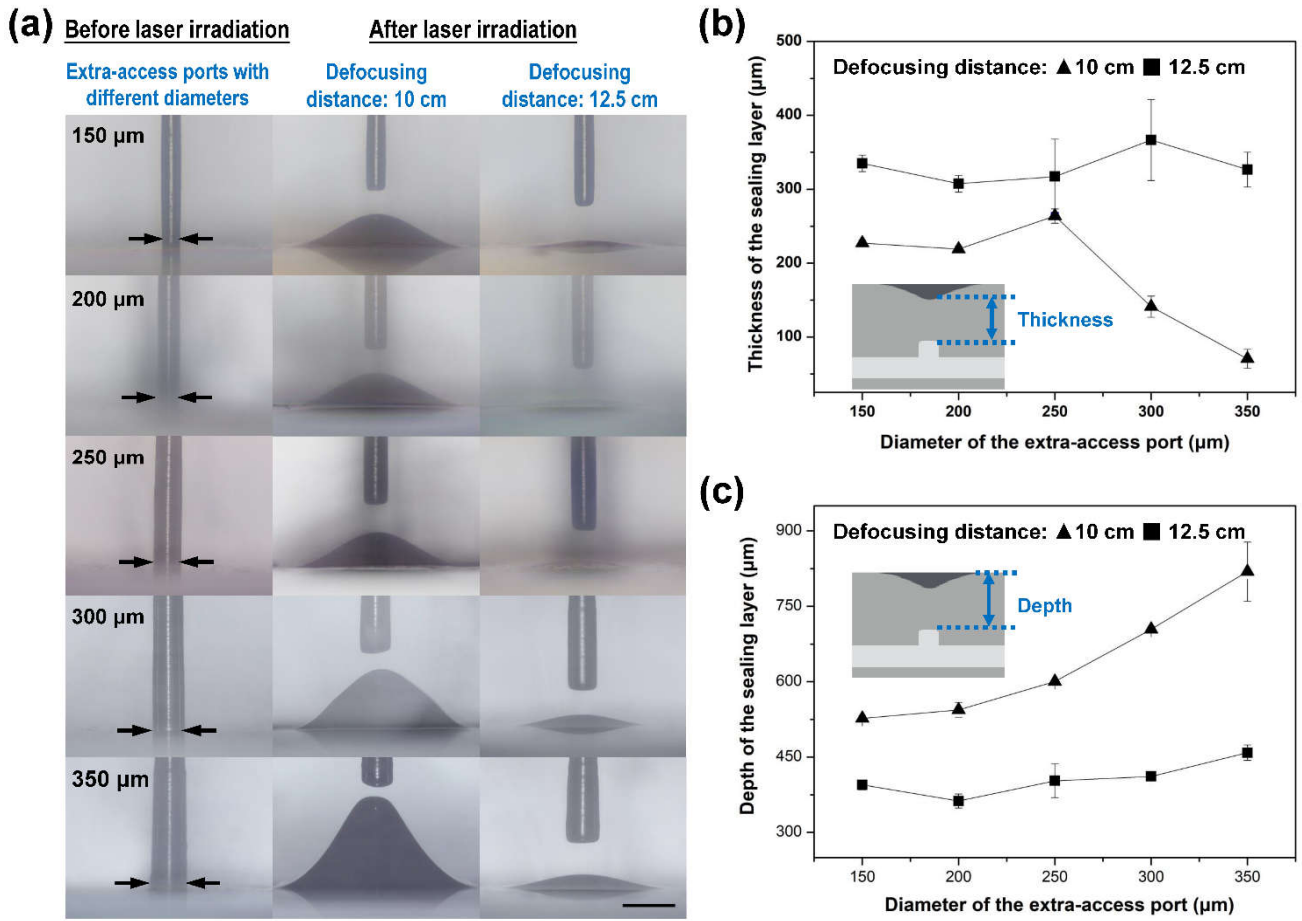

**Figure 2.** (a) Side-view optical micrographs of fabricated extra-access ports with different diameters ranging from 150 μm and 350 μm before and after $CO_2$ laser irradiation with defocusing distances of 10 and 12.5 cm. Average power and irradiation duration of the $CO_2$ laser beam were set at 33 W and 60 s, respectively. Scale bar: 500 μm. (b) Thickness and (c) depth of the sealing layer versus diameter of the extra-access port at defocusing distances of 10 and 12.5 cm.

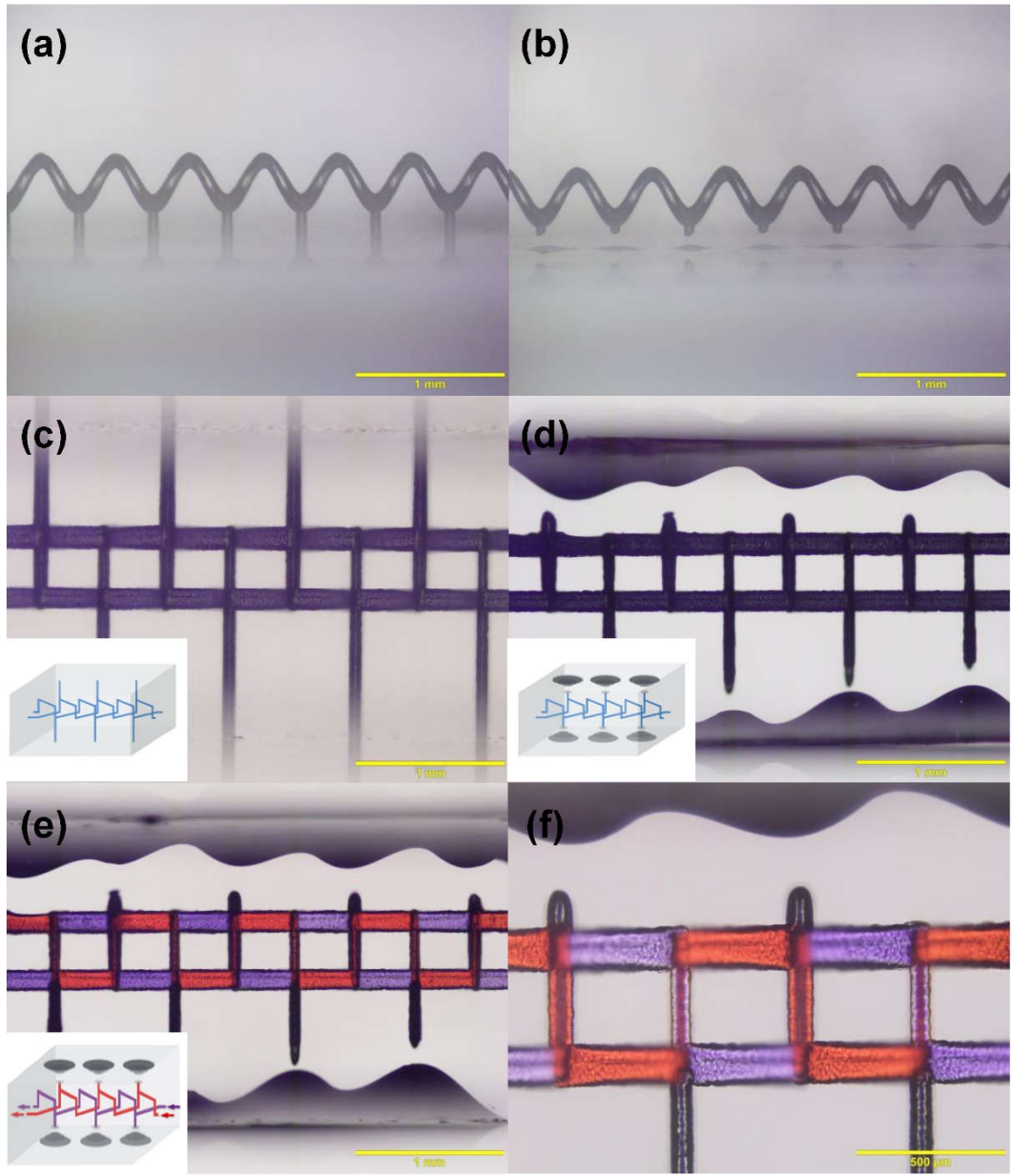

**Figure 3.** Optical micrographs of a 3D helical microchannel with a set of extra-access ports (a) before and (b) after $CO_2$ laser irradiation. Side-view optical micrographs of two 3D space-cross isolated microchannels with extra-access ports (c) before and (d) after $CO_2$ laser irradiation. Insets in (c) and (d) show 3D schematics of the microchannels before and after sealing, respectively. (e) Side-view optical micrographs of 3D space-cross isolated microchannels with sealed extra-access ports. Both the channels have been filled two different color solutions (red and purple). Inset in (e) shows 3D schematic of the sealed microchannels filled with the solutions. (f) is a close-up image of (e).

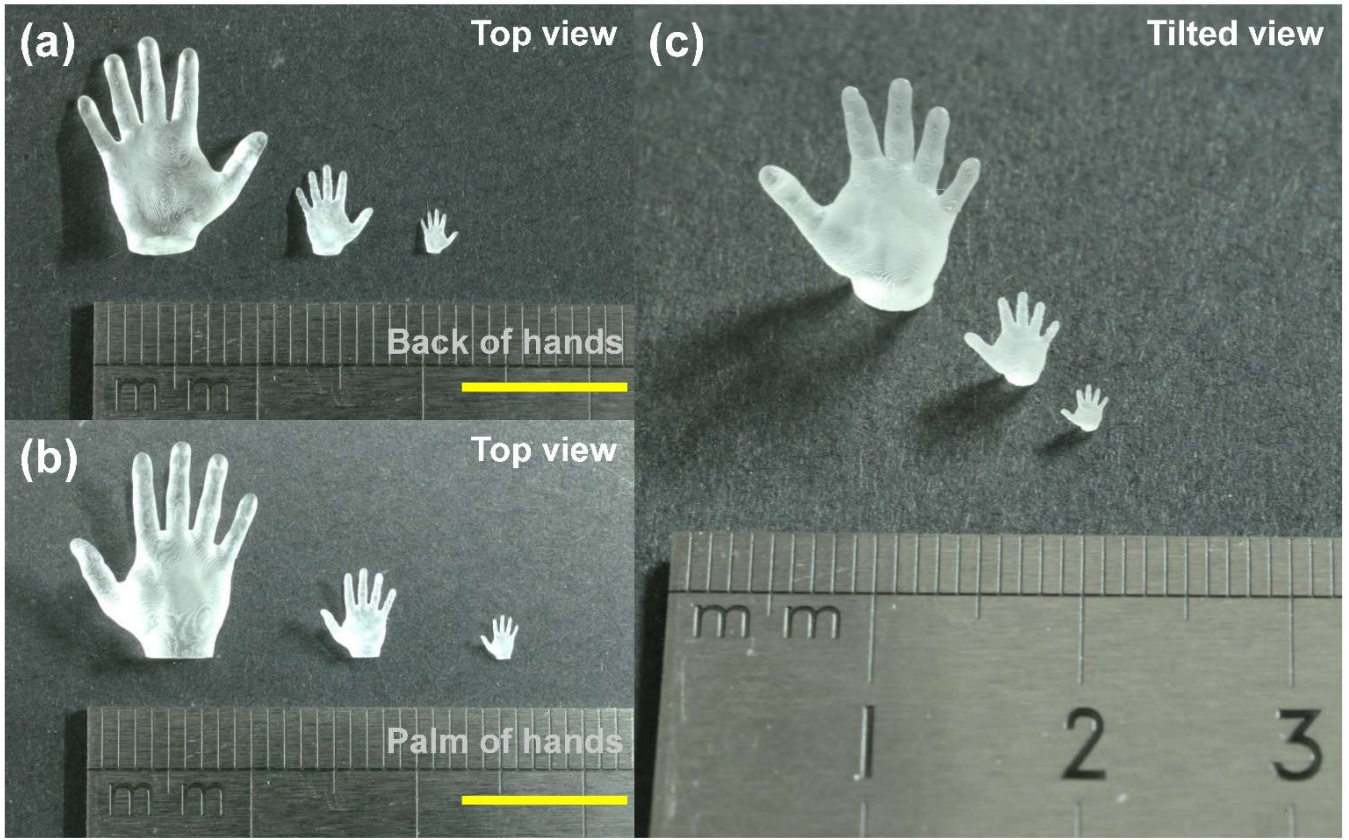

**Figure 4.** Ultrashort pulse laser subtractive printing of 3D glass hands with different sizes. Top-view photos of (a) the back and (b) the palm of three glass hands with scalable footprint sizes (left: ~1 cm × 1.2 cm × 0.4 cm, middle: ~0.5 cm × 0.6 cm × 0.2 cm, right: ~0.25 cm × 0.3 cm × 0.1 cm). Scale bar in (a-b): 1 cm. (c) Titled-view photo of the glass hands. Laser pulse duration was 4 ps.

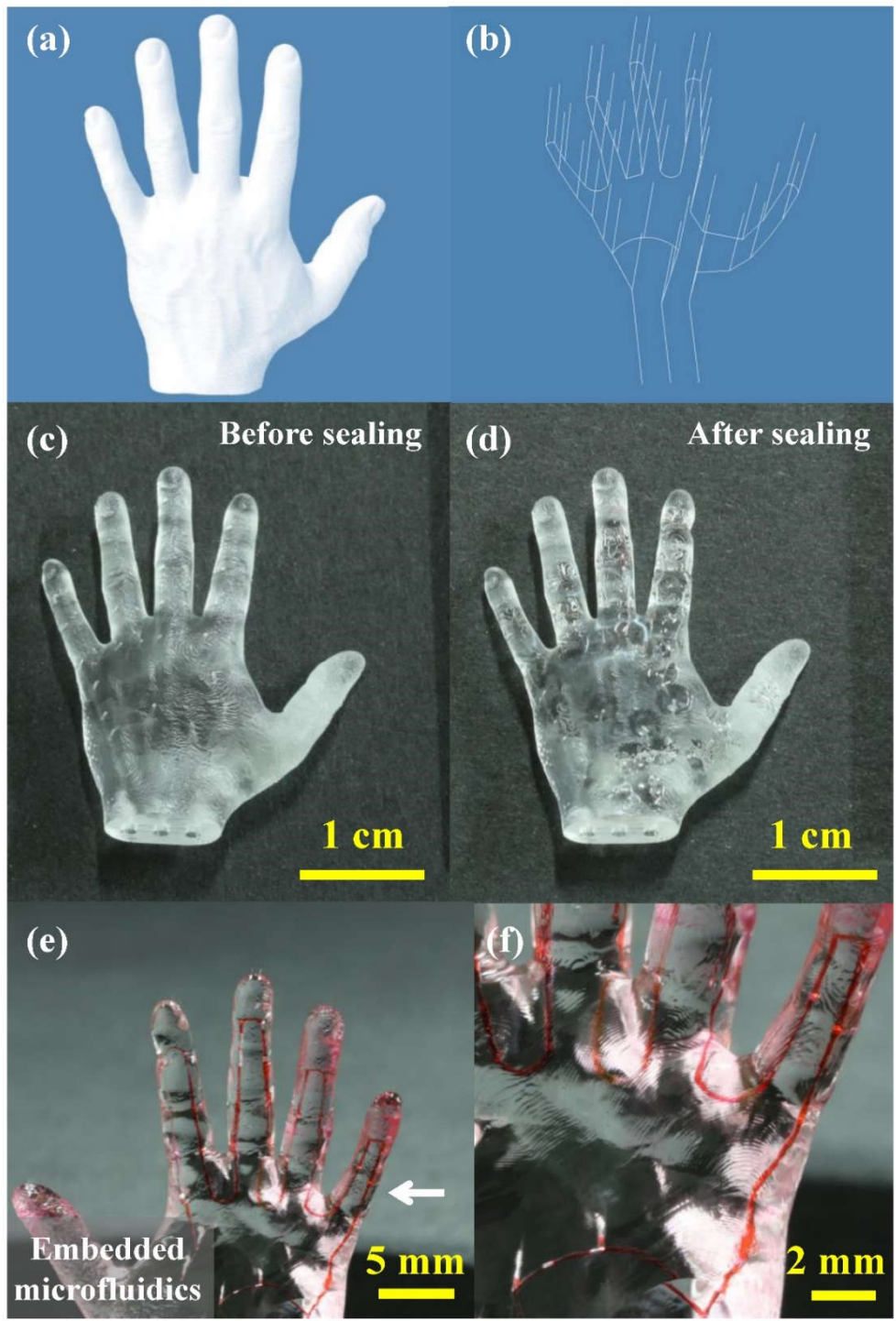

**Figure 5.** Schematics of (a) a 3D hand model and (b) an embedded microchannel pattern for ultrashort pulse laser assisted chemical etching. Pictures of 3D printed bioinspired glass hand with encapsulated channels and extra-access ports (c) before and (d) after $CO_2$ laser sealing. (e) 3D microfluidic demonstration of the hand based on combination of 3D glass subtractive printing and hybrid laser microfabrication. (f) Close-up image of a part of the hand as indicated by a white arrow in (e).

27